\documentclass[prl,preprint]{revtex4}
\usepackage{epsfig,graphicx}
\begin{document}

\title{ Ballistic Intrinsic Spin-Hall Effect in HgTe Nanostructures}

\author{C. Br\"une$^{1}$, A. Roth$^{1}$, E.G.\ Novik$^{1}$, M. K\"onig$^{1}$, H. Buhmann$^{1*}$,  E.M. Hankiewicz$^{2}$, W. Hanke$^{2}$, J. Sinova$^{3}$, and L. W. Molenkamp$^{1}$}

\affiliation{$^{1}$ Physikalisches Institut (EP3), Universit\"at
W\"urzburg, 97074 W\"urzburg, Germany}

\affiliation{$^{2}$ Institut f\"ur Theoretische Physik und
Astrophysik, Universit\"at W\"urzburg, 97074 W\"urzburg, Germany}

\affiliation{$^{3}$ Department of Physics, Texas A\&M University,
College Station, USA}


\date{\today}

\begin{abstract}
We report the first electrical manipulation and detection of the mesoscopic intrinsic spin-Hall
effect (ISHE) in semiconductors through non-local electrical measurement
in nano-scale H-shaped structures built on high mobility HgTe/HgCdTe quantum wells.
By controlling the strength of the spin-orbit splittings and the $n$-type to $p$-type
transition by a top-gate, we observe a large non-local resistance signal due to the ISHE in the
$p$-regime, of the order of k$\Omega$, which is several orders of magnitude larger than in metals. In the
$n$-regime, as predicted by theory, the signal is at least an order of magnitude smaller.
We verify our experimental observation by quantum transport
calculations which show quantitative agreement with the experiments.
\\

\end{abstract}

\maketitle

\section{Introduction}
Control,  manipulation, and detection of spin polarized carriers are the
focal goals of spintronics \cite{Awschalom07}. The creation
of new technologies based on spin current manipulation requires
new methods and materials for generating and controlling
spin-based properties of active devices. While progress has been
made in spin injection from a ferromagnetic metal into a
semiconductor through tunneling barriers, its detection efficiency
is still problematic. Applications of ferromagnetic
semiconductors are challenged by their ferromagnetic
transition temperatures which remain below room temperature.
A clear avenue to circumvent several of these key problems
is the direct use of electric fields to manipulate electron spins through
spin-orbit coupling based effects in paramagnetic systems. Of these class of effects
one of the premier candidate at present is the spin-Hall effect (SHE)
\cite{Dyakonov71,Hirsch99,Hankiewicz06,Handbook,Murakami03,Sinova04}
in which a transverse spin accumulation is created when an
electric current is passed through a material with strong
spin-orbit coupling, coming either from the band structure
(intrinsic spin-Hall effect, ISHE) or from the scattering of
electrons on heavy impurities (extrinsic spin-Hall effect, ESHE). Although the
SHE has been very actively studied theoretically over the last few
years \cite{Handbook}, experimentally only a few results have been reported due to the difficulty of
detecting the effect. Early experiments demonstrating the effect utilized
sensitive optical techniques \cite{Kato04,Wunderlich05,Sih05}. Electrical detection
of the SHE, although much more desirable from the device point of view,
is still more challenging  and has been demonstrated only in
metallic nanostructures \cite{metalSHE,Weng08}
The  detected
signals are weak, to a large extent due to the fact that samples
are in the diffusive transport regime in mostly weak spin-orbit coupled systems. While most of the above
experiments appear to result from an ESHE, the authors  of
\cite{Wunderlich05} and \cite{Weng08} have attributed their observations the ISHE, involving the
actual band structure spin-orbit coupling effects. A larger effect can be expected in
samples where the transport is ballistic and unequivocally stems from
the ISHE \cite{Hankiewicz04,Nikolic04}. Such an experiment is the
topic of this paper: we demonstrate electrical detection of the ballistic ISHE
in high mobility HgTe-based nanostructures.

\section{Experimental setup}

\begin{figure}
\centering
\includegraphics[width = 16cm]{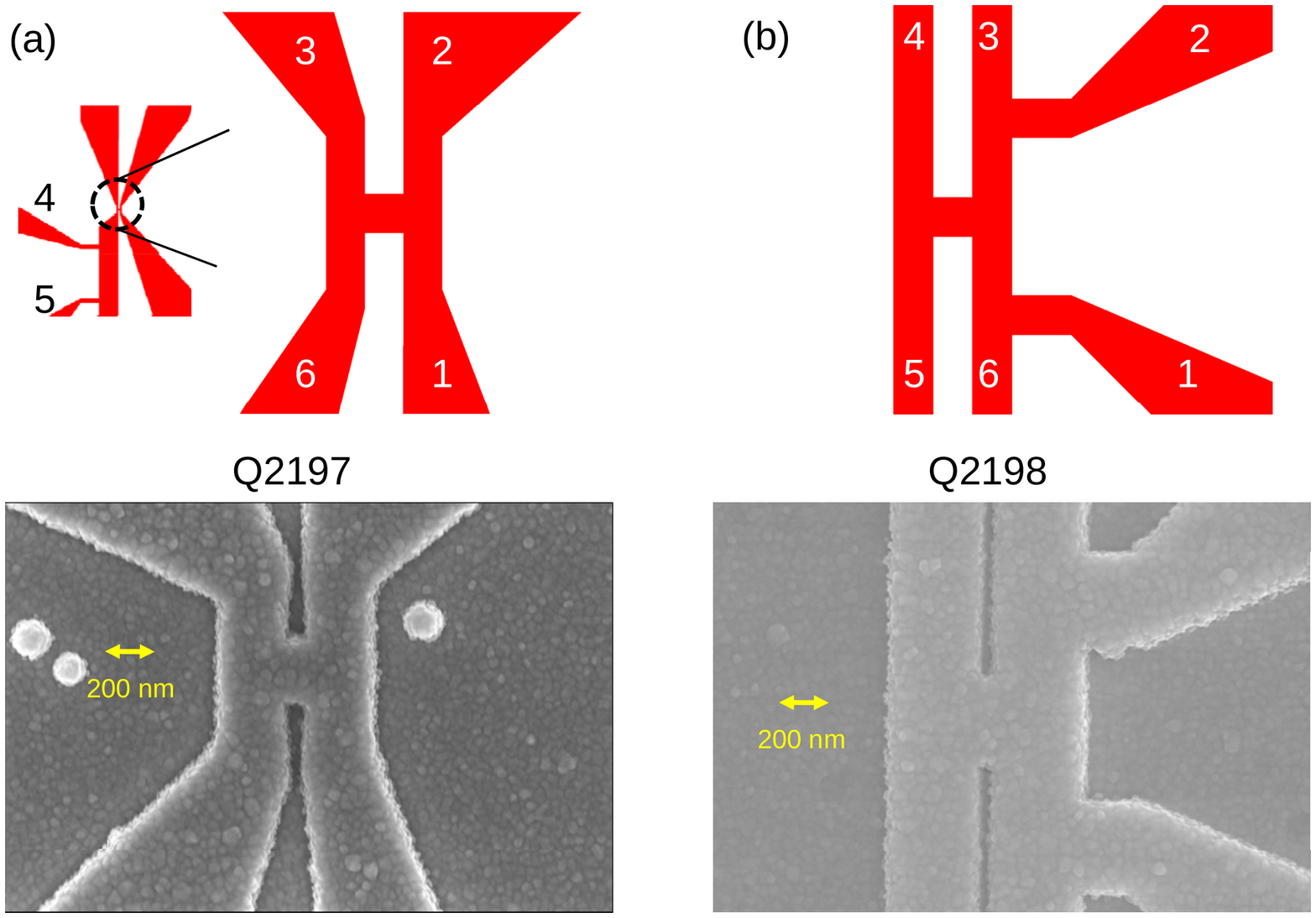}
\caption{Sample layout and electron beam micrograph of samples (a)
Q2197 and (b) Q2198.}\label{fig1}
\end{figure}

HgTe is a zero gap semiconductor which forms an inverted  type-III
QW with Hg$_{0.3}$Cd$_{0.7}$Te barriers when the well width is larger than
6.3 nm \cite{Koenig07}. The two-dimensional electron gas (2DEG) in
these structures exhibits a high carrier mobility (up to
$7\times10^5$ cm$^2$/Vs \cite{Becker07}) and a large,
gate-controllable, Rashba-type spin-orbit splitting of up to 30 meV
\cite{Gui04,Hinz06}. The external top-gate also allows to
change the carrier density from n-type through insulating to p-type
regime. The samples studied in this paper contain a 8 nm wide
symmetrically (Q2197) or asymmetrically (Q2198) $n$-doped HgTe QW.
The spacer width is 10 nm. A mean carrier mobility of $\mu = 2.5
\times 10^5$ cm$^2$/Vs ($\mu = 1.2 \times 10^5$ cm$^2$/Vs) was
deduced for an electron density of $n_s = 1.7\times 10^{11}$
cm$^{-2}$ ($n_s = 2.0\times 10^{11}$ cm$^{-2}$) on an ungated
Hall-bar of sample Q2197 (Q2198). From Shubnikov-de Haas
measurements, we estimate that the mobility is ten times smaller
in the p-type regime, which is mainly due to the larger effective mass.

In order to electrically detect the ISHE, we have fabricated H-shaped
mesa structures (see Fig.~1) using electron beam lithography and
dry-etching techniques, following a design proposed previously by some of us
in Ref.~\cite{Hankiewicz04}. The Au/Ti electrode is deposited on top of a 110 nm thick
SiO/SiN gate insulator layer which covers the entire sample. Ohmic
contacts are fabricated by thermal In-bonding.  Two additional
contacts have been added to the H-structures to allow further
characterization measurements. These contacts are attached to one
leg either far away from (sample Q2197, Fig.~1a) or in close
proximity to the H-bar (sample Q2198, Fig.~1b). The
H-structures consist of legs 1 $\mu$m long and
200~nm wide, while the connecting part is 200 nm
wide and 200 nm long. [Note that the micrographs in Fig. 1 show the top gate,
which masks the exact dimensions of the underlying H-mesa. Actually dimensions were verified from
micrographs of adjacent mesa structures taken before gate deposition.]
The estimated mean free path in these systems $l\geq 2.5 \; \mu{\rm m}$
which establishes that the samples studied are well within the quasi-ballistic regime.

The idea of the transport measurements is as follows \cite{Hankiewicz04}. When an
electric current flows in one of the legs of the H-bar structure
(say between contacts 1 and 2 in Fig.~1a), a transverse spin
current due to the intrinsic spin-Hall effect is induced in the
connecting part. Subsequently, this spin current
produces, due to the inverse spin-Hall effect
\cite{Hankiewicz05}, a voltage difference in the opposite leg of the H-bar structure
(in our example, between contacts 3 and 6) which can be measured by a
voltmeter \cite{Hirsch99,Hankiewicz04}. The H-shape of the
structure allows for a strong suppression of the residual voltage between contacts 3 and 6 that is
directly due to the potential difference between contacts 1 and 2 and the actual ISHE signal can be
easily identified by its dependence on the gate voltage.
Experimentally, all measurements were performed using standard AC lock-in techniques, using an excitation
voltage of 100 $\mu$V at a sample temperature of 1.8 K.

\section{Experimental results}

\begin{figure}
\centering
\includegraphics[width = 16cm]{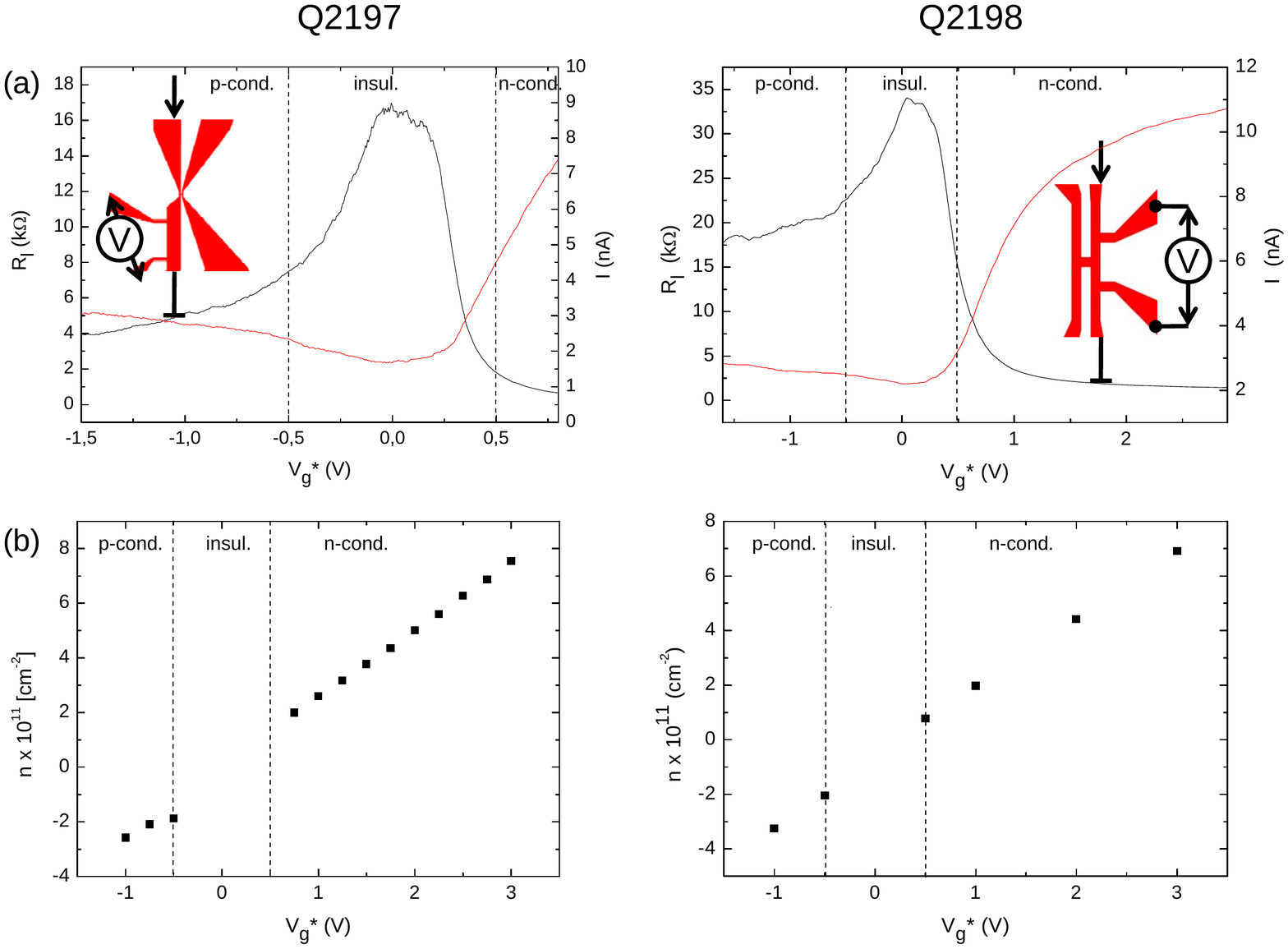}
\caption{(a) Signatures of the QSHE in measurement
configurations on one leg of the H-bar and (b) the dependence of the carrier density on the
applied gate-voltage. The contact configurations which lead to the
observation of the QSHE are indicated in the insets.} \label{fig2}
\end{figure}

Fig.~\ref{fig2} (a) shows the gate voltage dependence of the
sample current and the longitudinal resistance of samples Q2197
and Q2198. For reasons of comparison, we shift the gate voltage axis
such that renormalized gate voltage $V_g^* = 0 \;$V corresponds to the situation where
the bulk Fermi level is in the center of the energy gap
( the actual voltages where $V_g^*$ = 0 V were $V_g \approx -1.0 \; (-0.7)$ V for Q2197 (Q2198), respectively).
Fig.\ref{fig2} (b) shows the carrier density as a function of the
renormalized gate voltage. This data was obtained through
Hall-measurements on large Hall-bars fabricated from the same wafer
and demonstrates that we can vary the carrier concentration enough to
tune the sample from strongly $n$-type ($n \approx 8 \times 10^{11}$ cm$^{-2}$),
through the gap, down to a $p$-type regime at $V_g^{*} = -1$ V ($p \approx 3 \times 10^{11}$
cm$^{-2}$). For  $V_g^*$ between -0.5 V and 0.5 V the samples are insulating
(see Fig.~\ref{fig2}).

We recently showed \cite{Koenig07} that the inverted band structure
in HgTe quantum wells gives rise to the occurrence of the
quantum spin Hall effect(QSHE), a novel type of quantum Hall effect
that occurs at zero magnetic field, when the bulk of the sample is in the insulating regime.
In samples that are smaller than the
inelastic scattering length, the electrical conductance is then quantized
at $2e^2/h$. Both the current and the resistance data on Q2197
and Q2198 [cf. Fig.~\ref{fig2}~(a)] show indeed a resistance close to the
conductance quantum when the gate voltage tunes the sample into the insulating
regime ($-0.5~\rm{V}<V_g^*<0.5$~V) [the deviation in Q2198 comes
from the long voltage leads, see below].
As in the quantum Hall effect, QSHE quantization is caused
by the formation of one-dimensional edge channels. The non-local character of
carrier transport through these edge channels implies that the effect,
which is much stronger than the signal anticipated for the ISHE in our samples,
should also show up in our H-bar geometry. We have indeed observed
very strong non-local QSHE signals in the course of our experiments
on H-bars, and an extensive report of these results will be published elsewhere.
For our present objective of an observation of the ISHE, however, the QSHE is an
unwanted effect since it tends to swamp the ISHE signal - even though it shows
its maximum at a different gate voltage.

\begin{figure}
\centering
\includegraphics[width = 16cm]{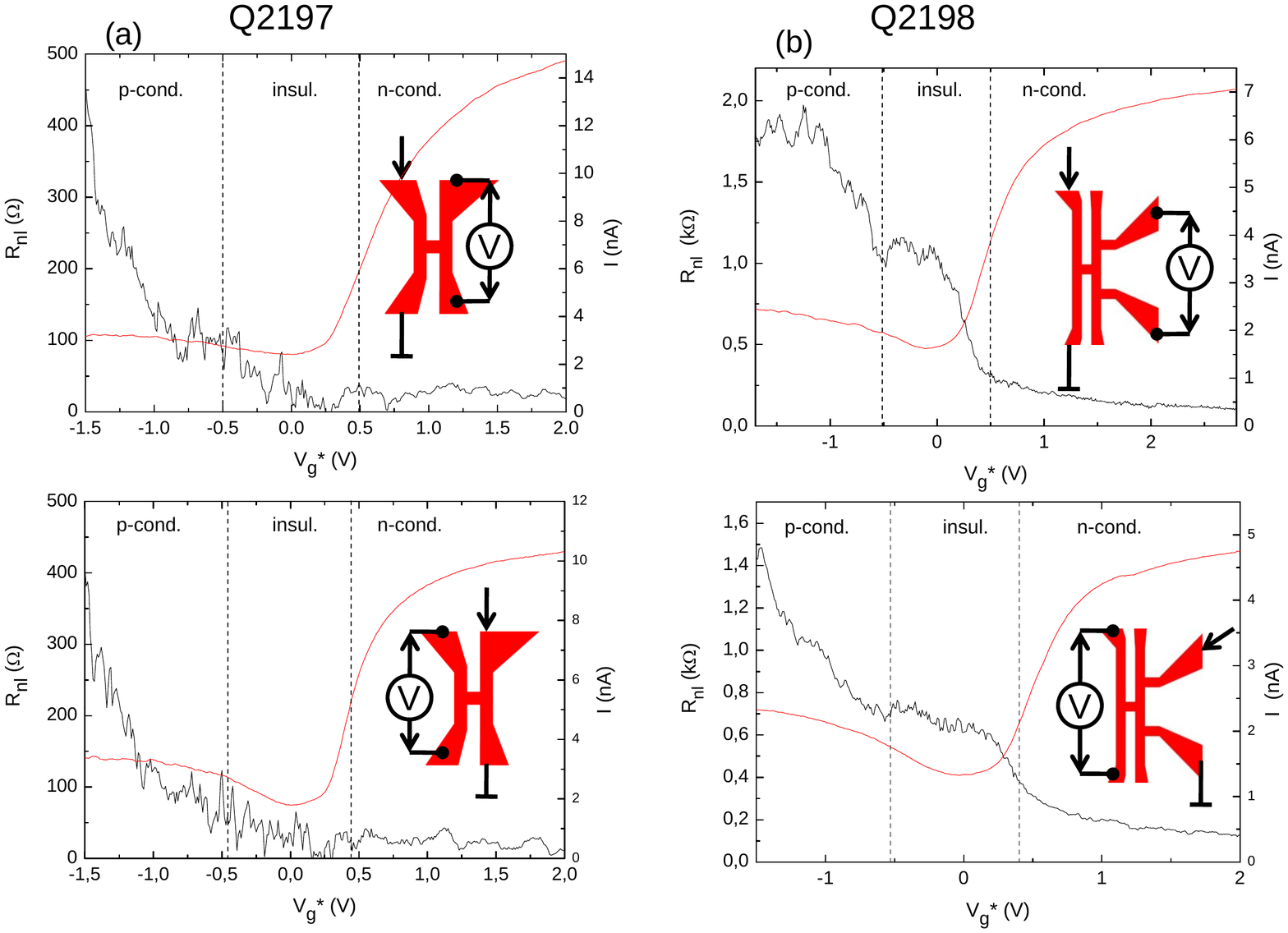}
\caption{Non-local resistance signal (black) and sample current (red) for (a) Q2197
and (b) Q2198 . The inset indicates the measurement configuration
for current injection (arrows) and voltage probes (V). The higher
current in the $n$-type regime follows primarily from the difference in
effective masses of the two regimes.} \label{fig3}
\end{figure}

The experiments shown in Figs.~\ref{fig3} represent two different approaches to suppress
the non-local QSHE signal. One approach is to make the devices sufficiently
small so as to provoke backscattering of the QSHE edge channels. Zhou et al.
recently showed theoretically~\cite{Niu08} that QSHE backscattering occurs
when the wave functions for opposite spin channels overlap, and estimate that this
happens for a device width of around 200-250~nm. This is the reason we report here
on very narrow devices. The second method to suppress the QSHE
is to choose a non-local configuration that implies edge channel transport over
distances (much) longer than the inelastic length. This is the reason we have
included the extra contacts on to the H-bar in Q2198: when the sample is insulating, the
extra contacts force the edge channels to take a detour before entering the horizontal part
of the H-bar.

Figs.~\ref{fig3} (a) and (b) show the current and non-local
resistance in a configuration suitable for picking up the ISHE signal
as a function of applied gate voltage for both samples. The upper and lower panels show the
results for an interchange of current and voltage contacts. One
observes that for both samples the non-local signal strongly
increases with gate voltage in the $p$-regime. The non-local
resistance is on the order of several 100~$\Omega$ for sample Q2197
and larger than a k$\Omega$ for sample Q2198. However, for the $n$-type regime
the signal is very low, does not show an appreciable dependence on
 gate voltage and thus can hardly be attributed to the ISHE.
For sample Q2197, any non-local contribution of the QSHE edge channels
is totally  suppressed, so that we attribute the observed
signal to the ISHE. The finite non-local
resistance around $V_g^*=0$ for sample Q2198 is an indication that we
have some residual non-local QSHE signal, possibly due to  a slightly
larger width of the horizontal bar. However, in the metallic regime,
the significant increase of the non-local resistance signal can only be
induced by the ISHE. For both samples, the strong non-local signal
in the $p$-regime remains almost unchanged even if the current and voltage
contacts are exchanged.
This is another strong argument that we observe the intrinsic
spin-Hall effect.

\begin{figure}
\centering
\includegraphics[width = 14cm]{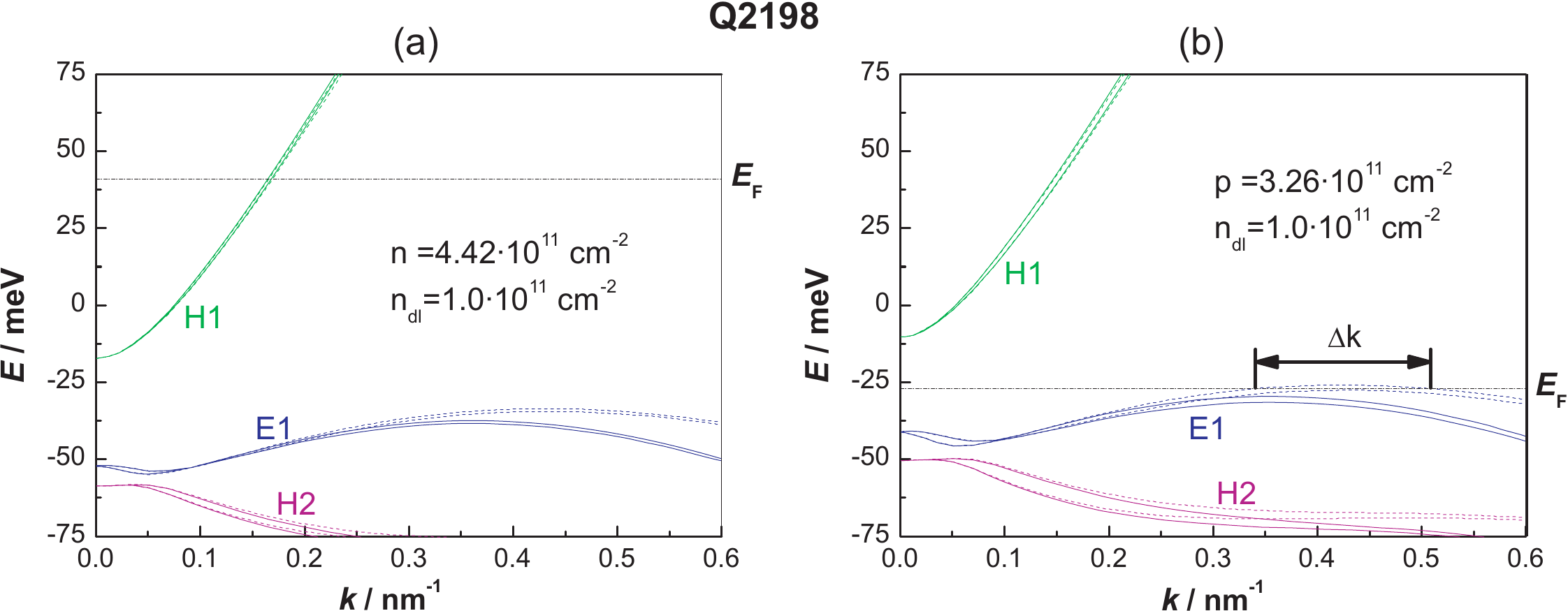}
\caption{ $8\times 8$ $\textbf{k}\cdot\textbf{p}$ band structure
calculation for the sample Q2198 in case of (a) $n$-type carrier
density $n = 4.42 \times 10^{11}$ cm$^{-2}$ and
(b) $p$-type carrier density of $p =3.26 \times 10^{11}$
cm$^{-2}$. The dashed-dotted line indicates Fermi level.
The band structure calculations for sample Q2197 look similar.}\label{fig4}
\end{figure}

\section{Theoretical simulations}

A key advantage of studying the ballistic ISHE is that methods to study the spin-dependent transport in this
regime, e.g. the Landauer-B\"uttiker (LB) formalism, are well established and can give
quantitative agreement with observed results, hence establishing a direct link between
experiment and theoretical expectations.
To model the specific devices studied in the experimental part, we first extract the effective masses and spin-orbit parameters
from a series of $8 \times 8$ $\textbf{k}\cdot\textbf{p}$ band
structure calculations \cite{Novik05}, where the influence of the top gate voltage is
 included in a self-consistent manner. We evaluate these parameters in the vicinity of the Fermi energy,
since at low temperatures, the quantum transport can  be described by
propagating modes at the Fermi energy. Figs.~4 (a) and (b) show
the calculated band structures for two representative carrier
densities $n = 4.42 \times 10^{11}$ cm$^{-2}$ and $p= 3.26\times
10^{11}$ cm$^{-2}$. For HgTe/HgCdTe quantum wells with an inverted
band structure in the vicinity of $k=0$, the conduction band (conventionally labeled H1 \cite{Novik05}) has heavy-hole character while the valence band (E1) has electron-like character. However, hybridization
of the energy states starts to play a role for $k\neq 0$ and the
spin-orbit splitting of both heavy hole and conduction bands must
be considered in the form of a combination of linear and cubic
terms. This is clearly relevant in our samples, because we find that the Fermi level passes
through the H1 band around $k=0.15$ nm$^{-1}$ for the n-type regime
and through the E1 band around $k=0.35$ nm$^{-1}$ for the p-type regime
[see Fig.~4]. We find that the following Hamiltonian matches all
important features emerging from the band structure calculations close to the
Fermi level:
\begin{eqnarray}\label{realH}
\hat{H}=\frac{\hat{p}^2}{2m^*}+\lambda_1(\hat{\sigma}_xp_y-\hat{\sigma}_yp_x)\nonumber\\
+\frac{i\lambda_2}{2\hbar^3}(\hat{p}^3_{-}\hat{\sigma}_{+}-\hat{p}_{+}^3\hat{\sigma}_{-})
+H_{dis}
\end{eqnarray}
where $m^* =0.305 \; m_e$ for the $p$-regime and $m^*=0.03 \; m_e$
for $n$-type samples. $\lambda_1 > 0$ and $\lambda_2 < 0$ are the
spin-orbit coupling parameters and in the $p$-regime
$|\lambda_2/\lambda_1| =2.9$ and $|\lambda_2/\lambda_1| =2.5$ for
samples Q2197 and Q2198, respectively. In $n$-regime
$|\lambda_2/\lambda_1| =3.8$, and $|\lambda_2/\lambda_1|=4.5$ for
samples Q2197 and Q2198, respectively. $H_{dis}$ is the disorder
potential. The size of the structures used for the theoretical
model is chosen in accordance with the actual dimensions of the experimental devices.
 In the experiments, the gate voltage causes changes in the
spin-orbit splitting as well as in carrier density and as a
consequence in the Fermi energy. We include this effect in the
simulations by changing the carrier density with gate voltage
according to the experimental data of
Fig.~\ref{fig2}~(b). The Fermi energies depend on the strength of the
spin-orbit interaction and the carrier densities. To perform real-space LB
calculations we rewrite the continuum effective mass
Hamiltonian in a tight-binding form (shown in detail in the
online appendix). In the tight-binding calculations, the disorder is
calculated by randomly selecting the on-site energies in the range
[$-W/2,W/2$], where $W=\hbar/\tau$, and $\tau$ is the transport
scattering time calculated from the effective masses and the experimental
mobility values. From the experimental mobility data, we find a corresponding
disorder strength $W=0.155$ meV. For this value of $W$,
ten averages over disorder configurations are sufficient to obtain convergent results.
The weak dependence of the observed effect on disorder in these materials is not
surprising given that the disorder induced by short range scattering gives rise to vanishing
vertex corrections \cite{Zhang07}.

We work in the linear response regime and the voltages on
different probes are found within the LB formalism using
boundary conditions where a charge current of 10 nA, for the setup of
structure Q2197, is driven  between contacts 1 and 2,
while between contacts 3 and 6 the electric current is zero. In this
configuration the non-local resistance signal is $R_{nl}=
V_{36}/I_{1,2}$. Details of calculations can be found in the online
appendix and in Ref. \cite{Hankiewicz04}.

\begin{figure}
\centering
\includegraphics[width = 14cm]{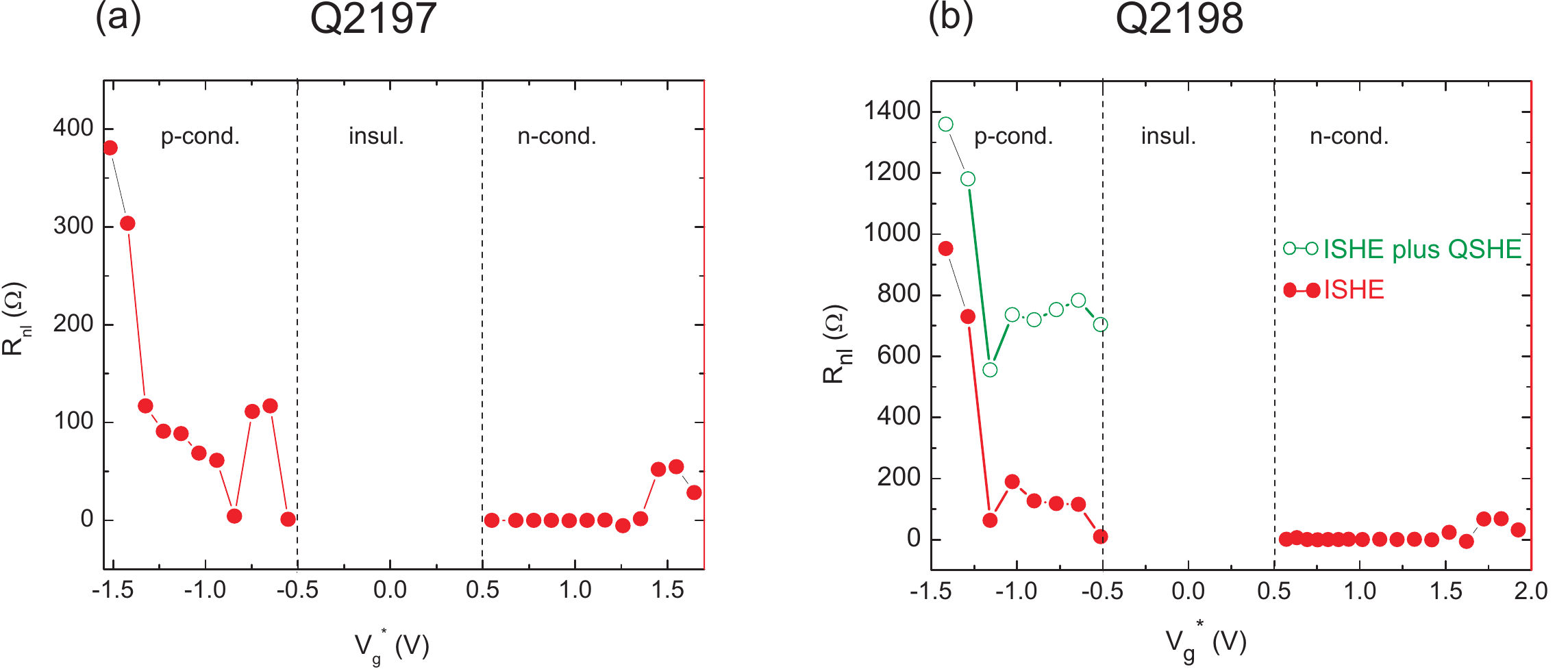}
\caption{Theoretical predictions of the resistance signal induced
by ISHE as a function of gate voltage for samples (a) Q2197 and
(b) Q2198. Close circles refer to the ISHE signal alone. In (b)
open circles include additionally the trace of QSHE extracted from
the experimental signal. For the sample Q2198, both curves overlap
in $n$-type regime. For both samples, experimental values of the
mobility and carrier densities were an input to the calculations.
The effective mass and strengths of spin-orbit coupling were
extracted from the band structure calculations (see text for
details).}
\end{figure}

Figs.~5 (a,b) show the theoretical predictions for the non-local
resistance signal as a function of gate voltage for samples Q2197 and
Q2198, respectively. For sample Q2198, we included the
theoretical prediction for the ISHE signal proper (closed circles) and the same, but adding
a signal due to the QSHE (open circles). This
QSHE signal has been taken directly from the experimental data,
assuming that if suppression of the QSHE signal were complete, the non-local
resistance signal should be zero in the insulating regime.
Therefore, Fig.~5 (a) and Fig.~5 (b) can be directly compared with
the experimental plots of Fig.~\ref{fig3}~(a) and (b),
respectively. Clearly, the theoretical results not only
show a very similar behavior as the experimental resistance signal,
but even have quantitative agreement. Furthermore, again in agreement with
the experiment, the theoretically predicted signals are at least an
order of magnitude stronger for the $p$- than for the $n$-regime.
As in the experiment, we also find in the calculations that Q2198
exhibits a stronger signal than Q2197. This stems from the fact that
for sample Q2198 the two extra voltage leads, used for suppressing the QSHE signal,
in the metallic regime correspond to two extra contacts situated very close to the
horizontal part of the H-bar.
The oscillating character of both experimental
and theoretical non-local resistance data stems from the fact
that the ratio of the Fermi energy to the spin-orbit splitting
changes over the range of gate voltage. In conclusion, our numerical
calculations are in a very good agreement with the experimental results
and confirm that the observed effect is indeed the ballistic ISHE.

\section{Summary}
In conclusion, we have presented experimental data evincing the first observation 
of a ballistic ISHE in HgTe nanostructures, as well as simulations that firmly confirm this interpretation.
We note that the effect measured is on order of a few \% of the excitation voltage, which is due to 
the fact that our signal is of second order in the spin-order coupling, in that the spin current induced by the
ISHE has to be turned into an electrical signal through the inverse SHE. Much larger effects can be
anticipated for an experiment that shows a signal that is directly proportional to the spin Hall current.
Such experiments are currently underway in our laboratory.

\acknowledgments
We thank J. Schneider for assistance in the experiments, and C. Gould, S.-C. Zhang and X.-L. Qi for stimulating discussions.
We gratefully acknowledge the financial support by the
German-Israeli Foundation (I-881-138.7/2005), ONR under grant
ONR-N000140610122, NSF under grant DMR-0547875, and SWAN-NRI. We
thank Leibniz Rechenzentrum M\"unich for providing computer
resources. J.S. is a Cottrell Scholar of the Research Foundation.


\end{document}